\documentclass[preprint,review]{elsarticle}

\journal{Neurocomputing}

\usepackage{amsmath}
\usepackage{amssymb}
\usepackage{graphicx}
\usepackage{subfigure}
\usepackage{epsfig}
\usepackage{epstopdf}
\usepackage{slashbox}
\usepackage{comment}
\usepackage{multirow}
\usepackage{xcolor,colortbl}
\usepackage{rotating}
\usepackage{algorithm}
\usepackage{algorithmic}

\newcommand{\RomanNumeralCaps}[1]{\MakeUppercase{\romannumeral #1}}
\usepackage{stackengine,scalerel}

\newcommand\overstar[1]{\ThisStyle{\ensurestackMath{%
			\setbox0=\hbox{$\SavedStyle#1$}%
			\stackengine{0pt}{\copy0}{\kern.2\ht0\smash{\SavedStyle*}}{O}{c}{F}{T}{S}}}}

\newcommand{\quotes}[1]{``#1''}

\begin{document}

\begin{frontmatter}

\title{Scalable Graph Neural Network-based framework for identifying critical nodes and links in Complex Networks}

\author[KSU]{Sai Munikoti\corref{CorrAuth}}
\author[CU]{Laya Das}
\author[KSU]{Balasubramaniam Natarajan}

\address[CU]{Complex Resilient Intelligent Systems Lab, Columbia University, New York, New York 10027, USA}
\address[KSU]{Department of Electrical and Computer Engineering, Kansas State University, Manhattan, Kansas 66506, USA}

\cortext[CorrAuth]{Corresponding author. Email:saimunikoti@ksu.edu}

\begin{abstract} Identifying critical nodes and links in graphs is a crucial task. These nodes/links typically represent critical elements/communication links that play a key role in a system's performance. However, a majority of the methods available in the literature on the identification of critical nodes/links are based on an iterative approach that explores each node/link of a graph at a time, repeating for all nodes/links in the graph. Such methods suffer from high computational complexity and the resulting analysis is also network-specific. To overcome these challenges, this article proposes a scalable and generic graph neural network (GNN) based framework for identifying critical nodes/links in large complex networks. The proposed framework defines a GNN based model that learns the node/link criticality score on a small representative subset of nodes/links. An appropriately trained model can be employed to predict the scores of unseen nodes/links in large graphs and consequently identify the most critical ones. The scalability of the framework is demonstrated through prediction of nodes/links scores in large scale synthetic and real-world networks. The proposed approach is fairly accurate in approximating the criticality scores and offers a significant computational advantage over conventional approaches.
\end{abstract}



\end{frontmatter}

\vspace{-0.2cm}
\section{Introduction}
Graph theory offers a powerful framework for studying the behavior of complex systems by representing a system as an interconnected network consisting of nodes and edges. Graph-theoretic techniques along with machine learning algorithms such as graph neural networks have recently gained popularity for studying both engineered and natural systems \cite{lu2020lstm,bongini2021molecular}. For instance, smart cities that are envisioned to connect billions of multi-modal sensors for optimized operation of the system are often modeled as graphs for analyzing interdependencies and robustness \cite{munikoti2021robustness}. Similarly, biological systems such as protein-protein interaction systems are typically represented and analyzed as graphs that allow a systematic study of a large number of collective biological behaviors and co-expressed features \cite{bongini2021molecular}. The interconnections in such systems represent interdependencies between different operating units that introduce interesting spatio-temporal dynamics in such systems.

Systems that are represented as graphs depend on the proper functioning of their constituent sub-systems (nodes) and their interconnections (links). However, typically there exists a set of \textit{critical nodes/links} that play a more crucial role in determining the output of the system than other (non-critical) nodes/links. These nodes/links represent a set of sub-systems and/or their interconnections, whose removal from the graph maximally disconnects the network, and thus severely disrupts the operation of the system. As a result, identification of critical nodes/links in complex networks is an important task for analysis and/or design of the underlying systems, and bears significance in several applications including social networks analysis, feature expression in biological networks, quality assurance and risk management in telecommunication networks, assurance of robustness in urban networks, control of social contagion, among others. For instance, identifying critical nodes/links in protein-protein interaction networks can be particularly useful in drug design. In such scenarios, critical nodes represent a minimum cardinality set of proteins whose removal would destroy the primary interactions and thus help neutralize potentially harmful organisms (e.g., bacteria or viruses) \cite{arulselvan2009detecting}. Similarly, in covert terrorist networks, critical nodes are individuals whose deletion from the graph will result in a maximum breakdown of communication between individuals in the network \cite{chaurasia2014use} and can significantly disrupt the operation of the network. In the case of engineered networks such as the internet, critical nodes/links are those which if compromised, will allow a hacker to increase the impact of a virus on the network, and which if strongly protected/shut down can efficiently stop the spread of a virus. Similar applications can be found in social networks, transportation engineering and emergency evacuation planning.

The problem of identifying critical nodes/links in a graph is becoming increasingly important across several domains. This is primarily because of the increase in the frequency of natural disasters as well as adversarial attacks in the recent past, causing large disruptions in engineered systems. For instance, there has been a surge in cyber attacks during the ongoing pandemic SARS-CoV-2 due to the inevitable rise in the usage of digital infrastructure \cite{pandey2020impact, kashif2020surge}. Similarly, in biological systems the use of high-throughput technologies to detect protein interactions has led to an exponential growth in the size of the protein interaction graphs in various species \cite{yang2020graph}. In such applications, the identification of critical nodes/links can assist in multiple ways such as equipping planner/operator to mitigate the impact of disruptions in infrastructure networks, easing the study of interactions in biological networks, etc.

As described earlier, critical nodes are those nodes whose removal maximally decreases the network functionality in terms of connectivity. This network functionality is often studied in the literature with the help of robustness metrics of the graph. Robustness quantifies the impact of loss of resources (nodes/links) on the performance of a system, and directly relates to the criticality scores of the nodes/links. There are numerous metrics and methodologies available in the literature that quantify the robustness of graphs such as effective graph resistance, flow robustness, total graph diversity, etc. \cite{alenazi2015comprehensive}. Owing to the inherent topological structure of graphs, each node/link contributes differently to the graph robustness, and hence its removal/loss affects the robustness to a different degree. In this regard, the notion of node/link criticality score is introduced, which quantifies the decrease in robustness (such as effective graph resistance) when the corresponding node/link is removed from the graph. Critical scores are then employed to rank the nodes/links, with the topmost rank assigned to the node/link whose removal maximally decreases the graph robustness and vice-versa. 

Several methods have been proposed to compute criticality scores based on graph robustness \cite{boginski2009identifying, wang2014improving, wang2015network}. However, such approaches typically measure the score of a node/link and repeat the process for all nodes/links in a graph. As a result of this iterative approach, such techniques exhibit computational complexity that increases drastically with the size of the graph, i.e., the number of nodes/links in the graph. For example, the complexity for identifying the optimal link whose removal maximally reduces robustness is of order $O(N^5)$ for a graph with $N$ nodes \cite{wang2014improving}. As a result, various efforts are directed towards approximating such algorithms and developing a computationally efficient solution \cite{van2011decreasing, wang2014improving}. However, with an increase in graph size, the accuracy of such approximations decreases, accompanied by a considerable increase in execution time. Moreover, several applications involve dynamically changing network topologies requiring node/link robustness to be dynamically estimated and maintained. The existing approaches being analytical in nature, and not exhibiting an inductive nature, one has to repeat the same procedure whenever the graph structure changes. Therefore, this paper proposes a graph neural network (GNN) based inductive learning approach to efficiently estimate the node/link criticality scores in complex networks. The criticality of an element being dependent on its neighborhood sub-graphs, the proposed framework leverages these sub-graphs to predict criticality scores.

\subsection{Related Work} 
The authors in \cite{wang2014improving}, explore effective graph resistance as a measure of robustness for complex networks. Robustness is improved by protecting the node/link whose removal maximally increases the effective graph resistance. The complexity of an exhaustive search to identify such a link is of order $O(N^5)$ for a graph of $N$ nodes. Therefore, multiple attempts are made in \cite{wang2014improving, van2011decreasing, wang2008algebraic}, where authors propose various node/link identification strategies. These approaches offer a trade-off between scalable computation and accurate identification of links. However, even with the proposed trade-offs, the lowest achieved complexity for the case of link identification is of order $O(N^2)$. To overcome computational complexity, the authors in \cite{pizzuti2018genetic}, propose a method based on a genetic algorithm to enhance network robustness. Particularly, the authors focus on identifying links whose removal would severely decrease the effective graph resistance of the graph. However, the algorithm requires fine-tuning of various parameters and is not scalable. 
There are some efforts of identifying critical nodes/links even in the case of critical infrastructure graphs. For instance, the authors in \cite{kermanshah2017robustness}, analyze the robustness of road networks to extreme flooding events. Here, they simulated extreme events by removing entire sections of a road system that correspond to nodes in the graph. They concluded that robustness in spatial systems is dependent on many factors, including the topological structure. However, the adopted method in this work is computationally exhaustive and it is difficult to infer anything about a new road network. In \cite{wang2015network}, the authors deploy effective graph resistance as a metric to relate the topology of a power grid to its robustness against cascading failures. Specifically, the authors propose various strategies to identify node pairs where the addition of links will optimize graph robustness. The minimum achieved complexity is of the order $O(N^2 -N + 2L_{c})$, where $L_{C}= \frac{N!}{2!(N-2)!-L}$ with $L$ links and $N$ nodes. Still, this computational complexity hinders its use in large complex networks. 

\subsection{Contributions}
With a growing emphasis on the diverse applications, there is a need to identify nodes/links in the graph whose removal significantly decreases the graph robustness. However, as seen in the previous discussion, existing methods to identify such nodes/links are computationally inefficient and do not adopt an inductive approach. Therefore, the key contributions of this work include:
\begin{itemize}
    \item A novel identification framework built on a GNN-based inductive learning approach is developed.
    \item The proposed framework is used to learn node/link criticality scores in a wide range of applications including biological, social and urban networks, and validated for efficient identification of critical nodes/links in large complex networks.
    \item The scalability of the proposed technique is demonstrated on synthetic and real-world networks by training GNN models on a small subset of nodes/links and predicting node/link scores in a relatively larger fraction of the graphs that were not used for training.
    \item The superior performance and computational efficiency of the technique are validated with existing methods for all graphs considered in the article. Results show that the mean accuracy of trained models is more than $90\%$ in identifying the Top-5\% of the critical nodes and links, and the execution time is three orders smaller compared to the conventional approaches. 
    \item The generalizability of the framework is demonstrated with the use of transfer learning, allowing new predictive models for other robustness metrics or alternate graph types (other than the ones used to train the GNN) to be obtained very efficiently from existing models.
\end{itemize}
The rest of the paper is organized as follows: Section \RomanNumeralCaps{2} describes various graph robustness metrics followed by a brief introduction of graph neural network. Section \RomanNumeralCaps{3} presents the proposed framework with experiments and results in Section \RomanNumeralCaps{4}. The final conclusions are provided in Section \RomanNumeralCaps{5}.

\section{Background}
This article proposes the use of graph neural networks (GNN) for the estimation of node/link criticality scores in a complex network. This is a two-step process, wherein the impact of a node/liink on the robustness of a graph is estimated by removal of the node/link, followed by a relative ranking/scoring of the nodes/links based on their impact. This involves the use of an appropriate metric of robustness, followed by the use of a GNN to capture the impact of nodes/links on the metric, and to perform relative ranking of the nodes/links. These components of the approach are discussed next.

\subsection{Metrics of Graph Robustness}
The robustness of a system is its ability to continue to function properly in the presence of disturbances/perturbations, such as failures of components. In a graph theoretic setting, this translates to the ability of the graph to function with loss of nodes/links. Several metrics have been proposed in the literature \cite{alenazi2015comprehensive} that quantify the robustness of a graph against random and targeted loss of nodes/links. The authors in \cite{alenazi2015comprehensive, alenazi2015evaluation}, extensively study various graph robustness metrics for different types of graphs and provide the most appropriate metrics for each type of graph. It has also been pointed out in the literature \cite{alenazi2015comprehensive} that there is no generic metric of robustness that can work for all types of graphs (i.e., with different degree distributions, average clustering coefficient, assortativity, etc.) under all scenarios (i.e, node/link removal, targeted and random removal, etc.). As a result, there exist several metrics depending on the objective and properties of the graph.

Effective graph resistance ($R_{g}$) is a widely used metric to quantify graph robustness \cite{wang2014improving, wang2015network, ellens2013graph} and is equal to the sum of the effective resistances over all the pairs of nodes \cite{ellens2011effective}. The effective resistance between any two nodes of a graph is computed by the commonly known series and parallel operations, by considering an electrical circuit-equivalent of the graph. EGR considers both the number of paths between nodes and their length (link weight), intuitively measuring the presence and quality of backup possibilities in the graph. The spectral form of $R_{g}$ can be expressed as:
\begin{equation}
    R_{g} = \frac{2}{N-1}\sum_{i=1}^{N-c}\frac{1}{\lambda_{i}},
    \label{eq:1}
\end{equation}
where, $\lambda_{i}$, $i=1,2,3,....N$ are the eigen values of the Laplacian matrix of a graph $G$ with $N$ nodes, and $c$ is the number of connected components in the graph. The expression in Equation (\ref{eq:1}) is a normalized expression, in that it allows the comparison of the metric, $R_{g}$ across graphs of different dimensions.

Weighted spectrum ($W_{s}$) is also a widely used metric for graph robustness \cite{fay2009weighted,long2014measuring}. It is defined as the normalized sum of $n$-cycles in a graph \cite{fay2009weighted}. An $n$-cycle in a graph $G$ is defined to be a sequence of nodes $u_{1}, u_{2},... u_{n}$ where $u_{i}$ is adjacent to $u_{i+1}$ for $i \epsilon [1, n-1]$  and $u_{n}$ is adjacent to $u_{1}$. It has been introduced to analyze the resiliency of internet topology. Thereafter, it has been compared to various other resiliency metrics and found to be a versatile metric especially for geographically correlated attacks \cite{alenazi2015evaluation}.
The $W_{s}$ can be expressed as,
\begin{equation}
    W_{s} = \sum_{i}(1-\lambda_{i})^n,
    \label{eq:2}
\end{equation}
where, different values of $n$ correspond to different graph properties. For instance, $n=3$ denotes the number of triangles in a graph with $W_{s}$ related to the weighted clustering coefficient. Similarly, with $N=4$, $W_{s}$ is proportional to the number of disjoint paths in a graph. Since this work focuses on quantifying the robustness of the graph based on connectivity, $n=4$ is used in this study. It must be noted here that although the work uses EGR and weighted spectrum as metrics of robustness, /the proposed approach is generic enough to be applied for any robustness metric.

\subsection{Graph Neural Networks}
Graph neural networks are a variant of artificial neural networks that are designed to capture patterns in data that can be represented in a graphical structure.
\cite{kipf2016semi} is one of the initial works that efficiently transforms the convolutional operations from Euclidean to graph space. The working principle of such models resembles that of convolutional neural networks and can work directly on graphs and exploit their topological information. The standard learning tasks on graph data are node classification, link prediction, graph classification, etc. A GNN typically involves learning node embedding vectors followed by feedforward layers for regression or classification tasks. The proposed learning algorithm in \cite{kipf2016semi}, depends on the size of the graph, which leads to scalability issues. To address this scalability issue, authors in \cite{hamilton2017inductive}, propose an inductive learning framework. Here, node embeddings are learned using subgraphs and thus the training is independent of graph size. Furthermore, this framework can be leveraged to infer about the unseen/new nodes of the graph belonging to the same family. The standard procedure to learn this embedding vector involves a \quotes{message passing mechanism}, where the information (node feature) is aggregated from the neighbors of a node and then combined with its own feature to generate a new feature vector. This process is repeated to generate the final embedding for each node of a graph. The GraphSAGE algorithm learns the mapping (aggregator) function instead of learning the embedding vectors. Hence, it can induce the embedding of a new node or node unseen during training, given its features and neighborhood. GNN addresses various learning tasks across domains including computer vision \cite{liu2020adaptive, yu2020resgnet}, natural language processing \cite{liang2021gated,yan2021quantum}, bio-chemistry \cite{bongini2021molecular}, etc. The proposed framework makes use of GraphSAGE as it is a state-of-the-art GNN modeling framework that is applicable to large graphs.

\section{Proposed Methodology}
This article proposes a two-step approach for inductive learning-based approximation of the criticality scores of nodes/links in a graph, referred to as Inductive Learner for Graph Robustness (ILGR).
The framework is first introduced for the identification of critical nodes, and then the approach to generalize the approach to links is elaborated. The first step involves the use of computationally manageable graphs to learn appropriate node embeddings and criticality scores with a GNN that allows faster learning of the criticality scores. This is followed by the prediction of criticality scores of nodes that have not been used for training the neural network. It must be noted, however, that in order for the neural network to reliably predict the scores at the time of deployment, the training and testing data, i.e., the nodes and the interconnections exhibit similar properties. This implies that the properties of the graphs used at the time of training should be as close as possible to the ones used for testing. In order to ensure that the training and testing graphs exhibit similar properties, one can use standard (synthetic) graphs such as power-law graphs and power-law cluster graphs that are known to exhibit properties similar to most real-world networks. However, in certain applications where such synthetic networks with similar properties cannot be obtained, it is possible to use a subset of the nodes of the testing graph and adopt transfer learning to tune the parameters of an already trained neural network model.
\begin{figure*}
\centering
	\includegraphics[width=15.0cm, height=6.5cm]{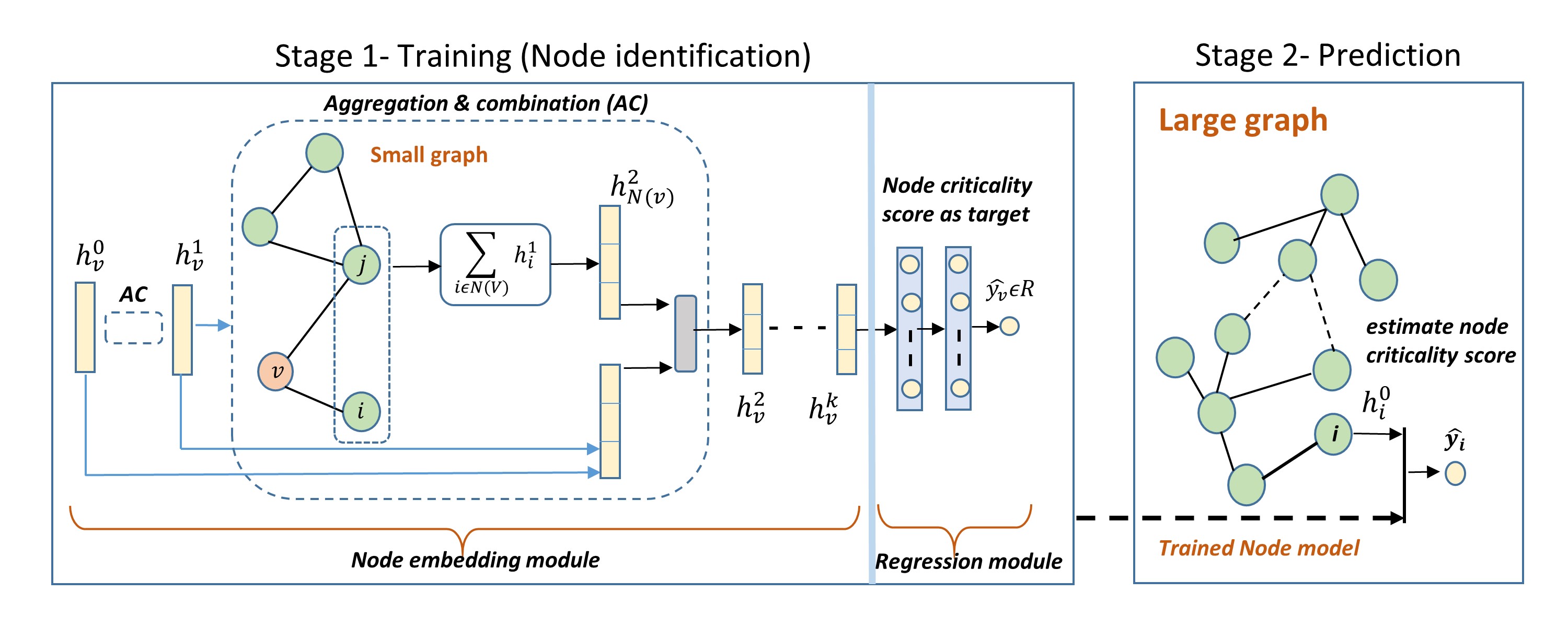}
	\caption{Proposed ILGR framework for node identification}
	\label{fig:1}
\end{figure*}

The proposed framework is illustrated in Figure \ref{fig:1}. The GNN model first learns \textit{node embeddings} from a set of features of nodes, which are then used to calculate the criticality scores. The following sections discuss in detail, the steps involved in obtaining the node criticality scores from a set of features, and neighborhood of nodes and training the models. 

\subsection{Node embedding module}
The first module of ILGR learns the embedding vector for each node by utilizing the graph structure and node target (criticality) scores. This is achieved in a manner such that nodes that are close in the graph space also lie close in the embedding space while maintaining a similar consistency between the node scores. As an initialization, the embedding of each node is composed of only the degree of the node, followed by a predefined number of ones. The node embeddings are learned based on GraphSAGE which is described briefly in the previous section. However, there are some modifications that have been made in the implementation. The Node embedding module is further subdivided into two tasks. The first task learns a representation for every node based on some combination of the representation of its neighboring nodes, parametrized by a quantity $K$, which quantifies the size of the neighborhood of nodes. Specifically, the parameter $K$ controls the number of hops to be considered in the neighborhood. For instance, if $K=2$, then all the nodes which are $2$ hops away from the selected node will be considered as neighbors. This defines the neighborhood of a node $v$ as:
\begin{equation}
N(v)=\{u:\text{D}(u,v)\leq K, \forall u\in G\}
\end{equation}
where $\text{D}(u,v)$ is a function that returns the the smallest distance between nodes $u$ and $v$. After defining the neighborhood, an aggregator function is employed to associate weights to each neighbor's embedding and create a neighborhood embedding for a selected node. Unlike previous works \cite{hamilton2017inductive, fan2019learning}, where weights are pre-defined, this work uses the attention mechanism to automatically learns the weights corresponding to each neighbor node as \cite{velivckovic2017graph}: 
\begin{equation}
    h_{N(v)}^{l} = \text{Attention}( Q^{l} h_{k}^{l-1} ) 
   \forall \hspace{0.1cm} k \epsilon N(v) 
   \label{eq:3}
\end{equation}
where, $h_{N(v)}^{l}$ represents the embedding of neighbourhood of a node $v$ in layer $l$ of the GNN, $h^{l-1}_{k}$ represents the embedding of $k^{th}$ neighboring node of $v$ in the $l-1$ layer of the GNN. Thereafter, for each neighborhood depth until $k=K$, a neighborhood embedding is generated with the aggregator function for each node and concatenated with the existing embedding for node $v$. However, the existing node embedding is not solely the output from the embedding of previous layer as implemented in various previous works. Rather, for existing node embedding, this article proposes to use the output of node embeddings from the previous two layers. This is similar to skip connections used by various researchers in the past for enhancing model performance for images and speech-related applications \cite{yamanaka2017fast, tu2017speech}, and can be expressed as:
\begin{equation}
    h_{v}^{l}= \text{Relu}(W^{l}[h_{v}^{l-1}||h_{v}^{l-2}||h_{N(v)}^{l}]) 
   \forall \hspace{0.1cm} k \epsilon N(v) 
   \label{eq:3a}
\end{equation}
where, $h_{v}^{l}$ represents the embedding of node $v$ in layer $l$ of the GNN, $h_{v}^{l-1}$ and $h_{v}^{l-2}$ denote the embedding of node $v$ in layers $l-1$ and $l-2$ respectively and $h_{N(v)}^{l}$ is the embedding aggregated from neighbors of $v$ as given in eqn. (\ref{eq:3}). This aggregation and combination process is  repeated for all layers of the model to obtain final node embeddings. The steps involved in the embedding module is summarized in Algorithm 1. $||$ in step $5$ of the Algorithm 1 is the concatenation operation.
\begin{algorithm}[h!]
 \caption{ ILGR embedding module}
 \begin{algorithmic}[1]
 \renewcommand{\algorithmicrequire}{\textbf{Input:}}
 \renewcommand{\algorithmicensure}{\textbf{Output:}}
 \REQUIRE Graph G, input node features $X_{v} \forall \hspace{0.1cm} v  \epsilon V$, unknown model weights $W$(combination weights) and $Q$(aggregation weights). 
 \ENSURE  Nodes embedding vector $z_{v}$ $\forall \hspace{0.1cm} v  \epsilon V$ . 
 \STATE Initialize: $h_{v}^{0}=X_{v}$  $\forall \hspace{0.1cm} v  \epsilon V$ 
  \FOR { layer $l=1$ to $l=L$ }
    \FOR { node $v=1$ to $v=V$ }
  \STATE  $h_{N(v)}^{l}$ = Attention($ Q^{l} h_{k}^{l-1} $) 
   $\forall \hspace{0.1cm} k \epsilon N(v) $ ;
  
  \STATE  $h_{v}^{l}=$ Relu($W^{l}[h_{v}^{l-1}||h_{v}^{l-2}||h_{N(v)}^{l}]$)   $\forall \hspace{0.1cm} v  \epsilon V$ ;
  \ENDFOR
  \ENDFOR
 \RETURN Final embedding vector $z_{v}=h_{v}^{L}$  $\forall \hspace{0.1cm} v  \epsilon V$ ;
 \end{algorithmic} 
 \end{algorithm}

\subsection{Regression module}
The output of the embedding module is passed through a regression module which is composed of multiple feedforward layers. The feedforward layers transform the embedding non-linearly and finally generate a scalar that denotes the node criticality score. The output of the $m^{th}$ layer in the regression module can be expressed as:
\begin{equation}
    y^{m} = f( W^{m}*y^{m-1} + b^{m} ) 
   \label{eq:4}
\end{equation}
where $W^m$ and $b^m$ represent the weights and biases in the $m^{th}$ layer, $f$ is the activation function such as ReLU, Softmax, etc. $y^{m}$ is the output of $m^{th}$ layer and $y^0=z_v$. The complete framework for node identification is depicted via Fig. \ref{fig:1}.

\begin{figure*}
\centering
	\includegraphics[width=15cm, height=4.5cm]{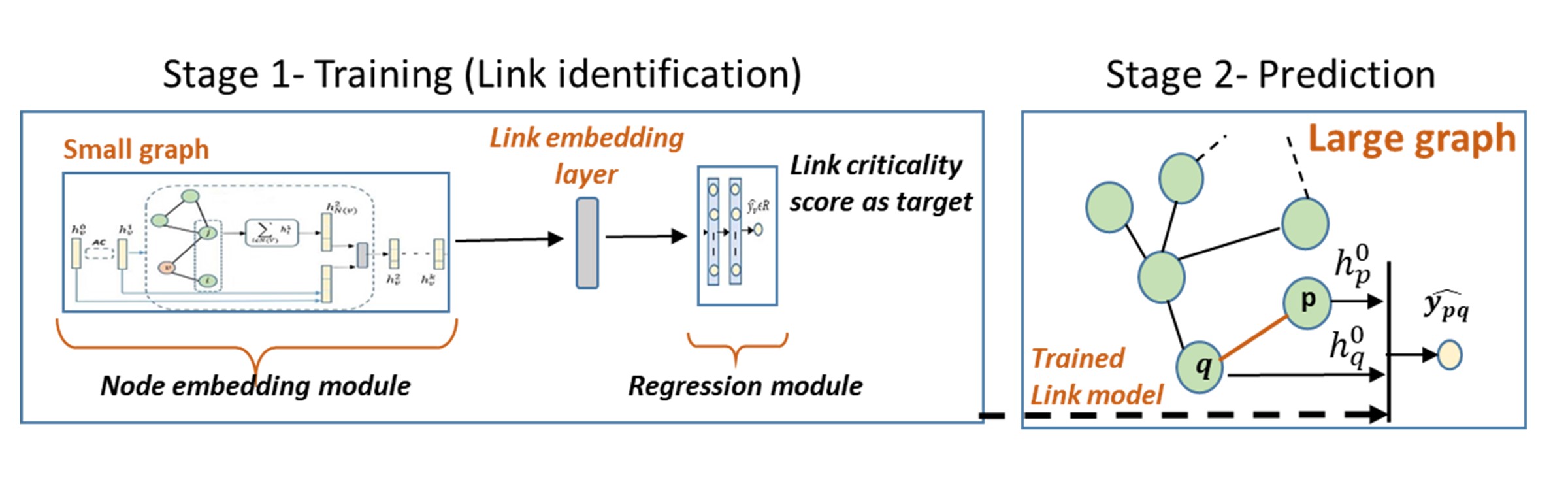}
	\caption{Proposed ILGR framework for link identification }
	\label{fig:2}
\end{figure*}

The framework for link criticality scores is almost similar to that of the node analysis with small differences. The output of the node embedding module is connected to a link embedding layer which generates link embedding from the associated pair of node embeddings. There are various ways to combine node embeddings including, concatenation, inner product, mean, L2 norm, etc. Here, the Hadamard product is used to generate link embedding. Thereafter, the link embedding is passed through the regression module to predict link criticality score. The entire framework of ILGR for link prediction is shown in Fig. \ref{fig:2}. 

The algorithm learns the weights of an aggregator function and feedforward layers by minimizing an appropriate loss function. In the proposed framework, the output of the model is node/link criticality score, which is further employed to rank nodes/links and identify the critical ones. So, instead of exactly learning the criticality values, it is sufficient to learn any real values provided the relative order of nodes remains intact. A suitable loss function for this scenario is the ranking loss \cite{chen2009ranking}. Unlike other loss functions, such as Cross-Entropy Loss or Mean Squared Error Loss, whose objective is to learn to predict a value or a set of values given an input, the objective of ranking loss is to preserve the relative distances between the inputs. Here, the pair-wise ranking loss has been used that looks at a pair of node ranks at a time. The goal of training the model is to minimize the number of inversions in the ranking, i.e., cases where a pair of node ranks is in the wrong order relative to the ground truth. This loss function can be expressed as:
\begin{equation}
    L_{ij}=-f(r_{ij})\log\sigma(\hat{y}_{ij})-(1-f(r_{ij}))\log(1-\sigma(\hat{y}_{ij}))
    \label{eq:5}
\end{equation}
where, $r_{i}$ is the ground truth value of criticality score for node $i$. $r_{ij}=r_{i}-r_{j}$ is the actual rank order, which the model is learning to infer through $y_{ij}=y_{i}-y_{j}$ by minimizing the loss $L_{ij}$. $f$ is a sigmoid function. The loss is aggregated for all the training node/link pairs, and then optimized to update the model weights. Once weights are learned, then an embedding vector and consequently the node/link scores can be predicted for a test node/link given its features and neighboring information. 
\begin{algorithm}[h!]
 \caption{ Algorithm of ILGR}
 \begin{algorithmic}[1]
 \renewcommand{\algorithmicrequire}{\textbf{Input:}}
 \renewcommand{\algorithmicensure}{\textbf{Output:}}
 \REQUIRE Model with unknown weights.
 \ENSURE  Trained model.  
 \STATE Generate ground truth criticality scores of nodes/links based on graph robustness score
  \FOR { each epoch}
  \STATE  Get each node embedding from embedding module. 
  \STATE  Estimate criticality score of nodes/links through regression module.
  \STATE  Update weights of both modules by solving eqn. (\ref{eq:3})
  \ENDFOR
 \STATE Predict node/link score on test graph.
 \RETURN Top $N\%$ of most critical nodes/links.
 \end{algorithmic} 
 \end{algorithm}

\subsection{Model settings and Training}
There are various hyper-parameters in the model that need to be tuned for training the model. The number of node embedding layers, i.e., the depth of the GNN is selected as three. The number of neurons in these layers are $64$, $32$, and $16$ respectively. The regression module consists of three feedforward layers with $12$, $8$ and $1$ neurons respectively. The activation function in all the layers is kept as \textit{relu}. The aggregation and combination operation involve trainable network weights. The ranking loss function is optimized via ADAM optimizer in the TensorFlow framework with its default settings. The training of both the embedding and regression modules is conducted end to end with input being a specific node/link along with its neighbor information and output being the corresponding criticality score.  

A different model is trained for each family of synthetic graphs. This is because, different families of graphs vary in their overall structure and link connections, i.e., degree distributions, assortativity, average clustering coefficient, etc. For a particular graph family, different random instances of graph are sampled, and then the ground truth of criticality scores are computed for each sampled graph using a conventional approach as stated in Algorithm 3. The model is then trained on multiple graphs of the same family. Thereafter, the trained model is tested on graphs of higher dimensions. For example, a model can be trained on multiple power-law graphs of dimension $100-1000$, and can be evaluated on power-law graphs of dimension $100-100000$. The training algorithm is summarized in Algorithm 2. The model is trained end to end on the Tensorflow framework. For real-world graphs such as US Power grid \cite{snapnets, nr}, Wiki-vote, already trained models are employed for predicting node ranks. Additionally, transfer learning is also implemented for a real-world network to demonstrate the efficacy of the framework. 
The experimental setup and results are shown in the next section. 

\section{Experimental Results}
This section discusses results obtained with the proposed framework for node and link criticality score prediction over a wide range of applications. The datasets used in this work to demonstrate the applicability of the proposed framework are first discussed, followed by the evaluation metrics used to report the performance of the framework. A baseline approach is used to compare the performance of the proposed framework with existing approaches.

\subsection{Datasets} We examine the performance of ILGR on both synthetic and real world graphs. The two commonly used synthetic graphs are generated using Python NetworkX library are as follows:
\begin{enumerate}
    \item \textbf{Power law}: Graphs whose degree distribution follow power law (i.e., heavy tailed), and many real networks have shown to be of this family \cite{barabasi2000scale}. It can be generated through a process of preferential attachment in which the probability that a new node $N_{y}$ connects with an existing node $N_{x}$ is proportional to the fraction of links connected to $N_{x}$.
    \item \textbf{Power law cluster}: Graphs which exhibit both power law degree distribution and clusters, and many real-world networks manifest these properties \cite{newman2000models}.
    As shown in \cite{holme2002growing}, one can construct a PLC graph by following a process of preferential attachment but in some fraction of cases ($p$), a new node $N_{y}$ connects to a random selection of the neighbors of the node to which $N_{y}$ last connected. 
    \end{enumerate}
    The real-world networks that are analyzed in this work are as follows \cite{hagberg2008exploring}:
    \begin{enumerate}
    \item \textbf{Bio-yeast}: It is a protein-protein interaction network for yeast consisting of $1458$ nodes and $1948$ links. \cite{newman2000models}.
    \item \textbf{US-powergrid}: It represents the western US power grid with $4941$ nodes representing buses and $6594$ links as transmission lines \cite{nr}.
    \item \textbf{Wiki-vote}: It contains all the Wikipedia voting data from the inception of Wikipedia till January 2008. $7115$ Nodes in the network represent wikipedia users with $103689$ links, where each link from node $i$ to node $j$ indicates that user $i$ voted a user $j$ \cite{snapnets}. 
    \item \textbf{cit-DBLP}: It is the citation network of DBLP, a database of scientific publications. There are $12591$ nodes and $49743$ links. Each node in the network is a publication, and each edge represents a citation of a publication by another publication \cite{nr}.
\end{enumerate}

\subsection{Evaluation metrics}
The trained model predicts the criticality scores of the test nodes/links in a graph, which are then used to identify the most critical nodes/links. In a general setting of robustness analysis, it is more relevant to identify top ranked nodes/links that are most critical, rather than knowing the ranks of all the nodes/links. Therefore, we have used Top-N\% accuracy to evaluate the proposed framework against the conventional approach. It is defined as the percentage of overlap between the Top-N\% nodes/links as predicted by the proposed method and the Top-N\% nodes/links as identified by conventional baseline approach, i.e., Algorithm 3 (discussed in next section). Top-N\% accuracy can be expressed as,

\begin{equation}
\small
   \frac{|\{\text{Predicted Top-N\% nodes/links}\} \cap \{\text{True Top-N\% nodes/links\}}|}{|V|\times (N/100)},
    \label{eq:4}
\end{equation}
where, $|V|$ is the number of nodes/links and $N$ is the desired band. In this work, the results are reported for Top-5\% accuracy. Further, the computational efficiency of the proposed approach is demonstrated in terms of execution time. More specifically, the execution time is same as wall-clock running time, i.e., the actual time the computer takes to process a program.
\begin{algorithm}[t]
 \caption{Conventional approach of identifying critical nodes/links on the basis of graph robustness}
 \begin{algorithmic}[1]
 \renewcommand{\algorithmicrequire}{\textbf{Input:}}
 \renewcommand{\algorithmicensure}{\textbf{Output:}}
 \REQUIRE Graph $G$ with $V$ nodes.
 \ENSURE  Node/link critical scores  
  \FOR { $n$  in \textit{V}}
  \STATE  Remove node/link $n$ from graph \textit{G} 
  \STATE  Compute robustness metric of the residual graph ($G-n$)
  \STATE  Assign criticality scores to node/link $n$
  \ENDFOR
 \STATE Rank nodes/links on the basis of above computed criticality scores. Top ranks correspond to more critical nodes/links.
 \RETURN Top $N\%$ of most critical nodes/links.
 \end{algorithmic} 
 \end{algorithm}
\subsection{Beseline approach}
To evaluate the performance of our proposed approach, we compare ILGR with a conventional method of estimating node/link criticality. The classical methodology involves an iterative method of removing a node/link from the graph and computing the robustness metric of the residual graph. The term \quotes{residual graph} referred to a leftover graph after the removal of node/link. This process repeats for all the nodes/links of the graph. Thereafter, the computed criticality scores of all the nodes/links are arranged to generate ranks and identify the most critical ones whose removal maximally decreases the graph robustness. Algorithm 3 summarizes the typical conventional approach to identify critical nodes based on graph robustness metrics.

\begin{figure}[h!]
\centering
	\includegraphics[width=6cm, height=3.5cm]{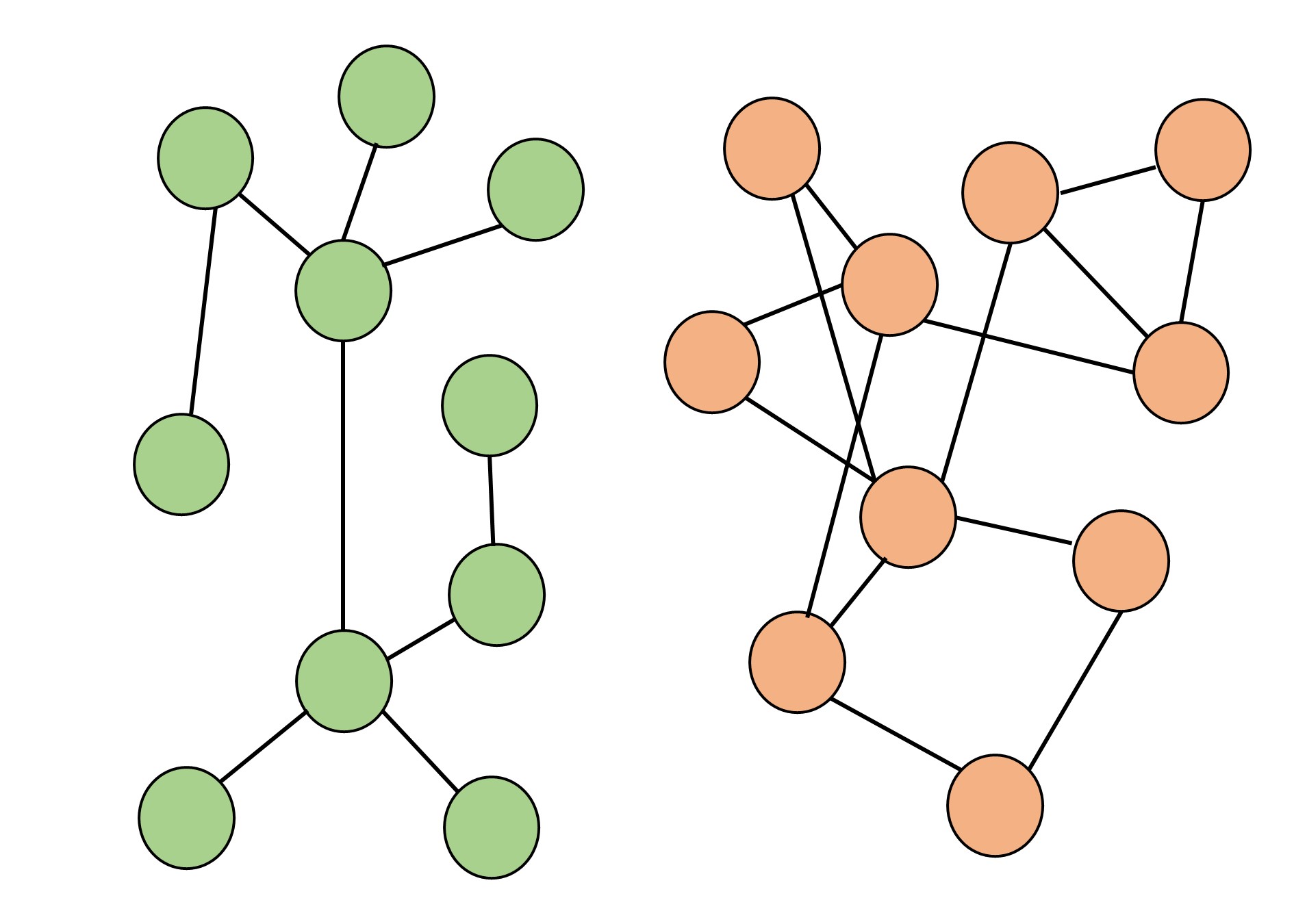}
	\caption{Left: Power law; Right: Power law cluster graph }
	\label{fig:3}
\end{figure}

\subsection{Results and Discussion for node identification}
This subsection access the performance of the proposed framework ILGR for nodes, in terms of Top-5\% accuracy and algorithm execution time. Different types of a graph with varying dimensions are considered for assessment. Table \ref{Table:Egr_syn_node} tabulates the Top-5\% accuracy of four different combinations of models, where the combination is generated based on graph family (i.e., power-law and power-law cluster) and robustness metrics (i.e., effective graph resistance ($R_g$) and weighted spectrum ($W_s$) ). The models are trained on $30$ graphs of dimension varying from $100$ to $1000$ nodes. The trained model is then employed to predict node criticality scores in power-law and power-law cluster graphs of higher dimensions, i.e., $500$, $1000$, $5000$, and $10000$ nodes.

\begin{table*}
    \centering
    \caption{Accuracy of node identification task in synthetic graphs.\\ PL: Power-law; PLC-Power-law cluster; Ntest: \# test nodes}
    \label{Table:Egr_syn_node}
	\begin{tabular}{|c|c|c|c|c|c|c|c|c|}
	\hline
    Degradation Level & \multicolumn{2}{c|}{Ntest 500} & \multicolumn{2}{c|}{Ntest 1000} & \multicolumn{2}{c|}{Ntest 5000} & \multicolumn{2}{c|}{Ntest 10000}   \\ 
    \cline{1-9}
    \backslashbox{graph}{scores} & Rg & Ws & Rg & Ws & Rg & Ws & Rg & Ws   \\ 
    \cline{1-9}
    \hline
    PL (100-1000) & 0.972 & 0.965 & 0.960 & 0.960 & 0.939 & 0.958 & 0.924 & 0.942 \\
    \hline
    PLC (100-1000)  & 0.943 & 0.952 & 0.937 & 0.950 & 0.935 & 0.948 & 0.919 & 0.930 \\
    \hline
  \end{tabular}
\end{table*}
It can be inferred from the Table \ref{Table:Egr_syn_node} that in the process of identifying Top-5\% of the most critical nodes, the mean accuracy of the PL model for robustness metrics $R_g$ and $W_s$ is $94.8$\% and $95.6$\%, respectively. The mean is taken across graphs of different node counts. Similarly, the accuracy of PLC model for $R_g$ and $W_s$ is $93.3$\% and $94.5$\%, respectively. The scalability of the framework is depicted via the model's high performance in graphs of increasing node dimensions. PL models perform relatively better than PLC models because the inherent topology is comparatively simpler in PL than that of PLC. More specifically, the learning mechanism of the ILGR is based on nodes sub-graphs, which generates embedding vectors by combining aggregation and combination operations. In PL graphs, there are very few nodes with a high degree while most of them bear small degrees. Therefore, node sub-graphs are more or less similar for most of the nodes. On the other hand, PLC graphs have a high clustering coefficient which introduces variability and complexity in sub-graphs, thereby makes the learning more challenging. This can be seen from the sample graphs in Fig. \ref{fig:3}. Nevertheless, our framework attains sufficiently high accuracy for both the models, which demonstrates the accuracy of the proposed framework in the task of estimating node criticality scores.
\begin{table*}
    \centering
    \caption{Accuracy of node identification task in real-world graphs. Ntest: \# test nodes}
    \label{Table:Egr_real_node}
	\begin{tabular}{|c|c|c|c|c|c|c|}
	\hline
    Robustness metric & \multicolumn{3}{c|}{Rg} & \multicolumn{3}{c|}{Ws}  \\ 
    \cline{1-7}
    \backslashbox{graph}{scores} & Ntest & Model & Top-5\% & Ntest & Model & Top-5\% \\
    \cline{1-7}
    \hline
    bio-yeast & 1500 & Rg-pl & 0.895 & 1500 & ws-pl & 0.877 \\
    \hline
    bio-yeast & 1500 & Rg-plc & 0.912 & 1500 & ws-plc & 0.898 \\
    \hline
    US powergrid & 4941 & Rg-pl & 0.86 & 4941 & ws-pl & 0.923 \\
    \hline
    US powergrid & 4941 & Rg-plc & 0.928 & 4941 & ws-plc & 0.914 \\
    \hline
    Wiki-vote & 7115 & Rg-pl & 0.865 & 7115 & ws-pl & 0.893 \\
    \hline
    Wiki-vote & 7115 & Rg-plc & 0.892 & 7115 & ws-plc & 0.887 \\
    \hline
    cit-DBLP & 12591 & Rg-pl & 0.875 & 7115 & ws-pl & 0.889 \\
    \hline
    cit-DBLP & 12591 & Rg-plc & 0.878 & 7115 & ws-plc & 0.895 \\
    \hline
  \end{tabular}
\end{table*}

The generalizability and the scalability of the proposed approach are further reinforced through Table \ref{Table:Egr_real_node}, which reports the Top-5\% accuracy in real-world networks, predicted through models trained on PL and PLC graphs. It can be observed that the PL model has a mean accuracy of $87.4$\% and $89.5$ \% for $R_g$ and $W_s$, respectively, where the mean is taken across four different real graphs. Similarly, the mean accuracy of PLC model is $90.0$\% and $89.8$ \% for $R_g$ and $W_s$, respectively. The proposed framework has sufficiently high accuracy in detecting critical nodes with both the robustness metrics, even though the model has never seen the real-world graph during the training period. This works because the nodes sub-graphs of real-world networks could match with that of synthetic graphs and therefore the model might have counter that type of sub-graphs during the training period. Furthermore, the models trained on PLC graphs perform better than that of PL graphs. The reason is that the PLC graphs manifest both power-law degree distribution and clusters, hence, are more accurate depiction of real-world networks compared to PL graphs. In addition, the model performance for any alternate graph family or robustness metrics can be enhanced via efficient re-tuning. In this regard, we implement transfer learning by tuning the model trained on PLC graph for estimating criticality scores in real-world bio-yeast network. Although we have used $150$ nodes for tuning, the model performance has been increased by $2.7$\% compared to the scenario when the model trained solely on PLC graphs is used for estimation. Thus, ILGR can accurately identify critical nodes for any large network in a very efficient manner which further strengthens its scalability.

Along with the accurate identification of nodes, the proposed framework provides an appreciable advantage in execution time which is indicated by the running times in Table \ref{Table:exe_time_node}. The proposed method is multiple orders faster than the conventional approach, and this gap will increase as the network size grows. All the training and experiments are conducted on a system with an Intel i9 processor running at $3.4$ GHz with $6$ GB Nvidia RTX $2070$ GPU. The time reported for the proposed approach only includes the prediction time as training is done offline. Even the training time of the proposed method is relatively less than that of the conventional approach.

\begin{table}[h!]
    \centering
    \caption{Running time of node identification task}
    \label{Table:exe_time_node}
	\begin{tabular}{|c|c|c|c|c|}
	\hline
    \backslashbox{graph}{specs} & Ntest & Model & Time: proposed (s) & Time: conventional(s) \\
    \hline
    PL & 5000 & Rg-pl &  17 & 64600 \\
    \hline
    PLC & 5000 & Rg-plc & 17 & 64600 \\
    \hline
    US Powergrid & 4941 & Rg-plc & 16 & 64212 \\
    \hline
    Wiki-Vote & 7115 & Rg-plc &  23 & 86420 \\
    \hline
  \end{tabular}
\end{table}

\subsection{Results and Discussion for link identification}

This subsection assesses the performance of the proposed ILGR framework in the task of critical link identification. The validation is done across different graph sizes and graph types. Separate models are trained for all possible combinations of graph type and metric type, which results in four distinct models. For each model, $30$ different random graphs of dimension varying from $1000$ to $2000$ links are generated for training. The ground truth for the selected links is computed with Algorithm 3. The training model is similar to that of node except that a layer is added that generates a link embedding vector from an associated pair of node embeddings. Consequently, the link embedding is stacked with a regression module to predict link criticality score. The trained model is then used to predict link scores in graphs of higher dimensions, i.e., $1000$, $2000$, $10000$, and $20000$ links. Table \ref{Table:Egr_syn_link} reports accuracy for $R_g$ and $W_s$ in PL and PLC graphs. It can be observed that the mean accuracy of PL model in detecting Top-5\% of the critical links is $91\%$ and $94\%$ for $R_g$ and $W_s$ both, respectively. Similarly, the mean accuracy of PLC model is $97.5\%$ and $96.1\%$ for $R_g$ and $W_s$, respectively. The model has fairly high identification accuracy even though the mean is taken across graphs of higher dimensions than that of training. There is a very nominal fall of accuracy with increasing size although the graph size scales in the order of two.
\begin{table*}
    \centering
    \caption{Accuracy of link identification task in synthetic graphs.}
    \label{Table:Egr_syn_link}
	\begin{tabular}{|c|c|c|c|c|c|c|c|c|}
	\hline
    Degradation Level & \multicolumn{2}{c|}{Ntest 1000} & \multicolumn{2}{c|}{Ntest 2000} & \multicolumn{2}{c|}{Ntest 10000} & \multicolumn{2}{c|}{Ntest 20000}   \\ 
    \cline{1-9}
    \backslashbox{graph}{scores} & Rg & Ws & Rg & Ws & Rg & Ws & Rg & Ws   \\ 
    \cline{1-9}
    \hline
    PL (200-2000) & 0.979 & 0.991 & 0.970 & 0.985 & 0.963 & 0.982 & 0.953 & 0.97 \\
    \hline
    PLC (200-2000)  & 0.981 & 0.968 & 0.98 & 0.961 & 0.972 & 0.96 & 0.967 & 0.957 \\
    \hline
  \end{tabular}
\end{table*}

The scalability and generalizability of our approach in the task of link identification are further supported by evaluating model performance on real-world networks. Table \ref{Table:Egr_real_link} tabulates the Top-5\% accuracy in four real-world networks. The accuracy is reported for all the four different models that have been trained on synthetic graphs. It can be inferred that the model has sufficiently high accuracy in identifying critical nodes for both robustness metrics, even though the model has never seen the real-world graph during the training period. Compared to PL graphs, models trained on PLC graphs seem to have high accuracy for real-world networks and the reason is the same as discussed earlier. Further, the execution times in Table \ref{Table:exe_time_link} demonstrate the computational efficiency of the proposed method. 
\begin{table*}
    \centering
    \caption{Accuracy of link identification task in real-world graphs. Ntest: \# test links}
    \label{Table:Egr_real_link}
	\begin{tabular}{|c|c|c|c|c|c|c|}
	\hline
    Robustness metric & \multicolumn{3}{c|}{Rg} & \multicolumn{3}{c|}{Ws}  \\ 
    \cline{1-7}
    \backslashbox{graph}{scores} & Ntest & Model & Top-5\% & Ntest & Model & Top-5\% \\
    \cline{1-7}
    \hline
    bio-yeast & 1948 & Rg-pl & 0.920 & 1948 & ws-pl & 0.904 \\
    \hline
    bio-yeast & 1948 & Rg-plc & 0.926 & 1948 & ws-plc & 0.952 \\
    \hline
    US powergrid & 6594 & Rg-pl & 0.946 & 6594 & ws-pl & 0.91 \\
    \hline
    US powergrid & 6594 & Rg-plc & 0.931 & 6594 & ws-plc & 0.946 \\
    \hline
    Wiki-vote & 30000 & Rg-pl & 0.887 & 30000 & ws-pl & 0.871 \\
    \hline
    Wiki-vote & 30000 & Rg-plc & 0.896 & 30000 & ws-plc & 0.925 \\
    \hline
    cit-DBLP & 49743 & Rg-pl & 0.870 & 49743 & ws-pl & 0.831 \\
    \hline
    cit-DBLP & 49743 & Rg-plc & 0.872 & 49743 & ws-plc & 0.892 \\
    \hline
  \end{tabular}
\end{table*}
\begin{table}[h!]
    \centering
    \caption{Running time of link identification task}
    \label{Table:exe_time_link}
	\begin{tabular}{|c|c|c|c|c|}
	\hline
    \backslashbox{graph}{specs} & Ntest & Model & Time: proposed (s) & Time: conventional (s) \\
    \hline
    PL & 5000 & Rg-pl &  19 & 64830 \\
    \hline
    PLC & 5000 & Rg-plc & 19 & 64830 \\
    \hline
    US Powergrid & 6594 & Rg-plc & 21 & 65470 \\
    \hline
    Wiki-Vote & 10000 & Rg-plc &  25 & 88231 \\
    \hline
  \end{tabular}
\end{table}

Although the performance of ILGR in the tasks of node identification and link identification from the estimated criticality scores are similar to a large extent ( due to common framework, i.e., node embeddings of sub-graphs), there are some differences. First of all, the overall performance of ILGR for link identification is better than that of node identification as seen from the Tables \ref{Table:Egr_real_node} to \ref{Table:Egr_real_link}. This is because, a link embedding vector is dependent on two nodes' embedding vectors, thereby including more information compared to the node case which solely uses single node embeddings. The second difference can be seen in their execution times. Link identification takes more time than that of a node as it involves an extra operation (i.e., the combination of node pair embeddings to generate link embedding) apart from common executions.

\section{Conclusions and Future work}

This paper proposes a graph neural network based ILGR framework for fast identification of critical nodes and links in large complex networks. Criticality score is defined based on two graph robustness metrics, i.e., effective graph resistance and weighted spectrum. ILGR framework consists of two parts, where in the first part, a graph neural network based embedding and regression model are trained end to end on synthetic graphs with a small subset of nodes/links. The second part deals with the prediction of scores for unseen nodes/links of the graph. The Top-5\% identification accuracy of the model is more than $90\%$ for both the robustness metrics. Further, the scalability of the model is shown by identifying critical nodes/links on real-world networks. The proposed approach is multiple orders faster compared to the conventional method. As part of future work, we will systematically incorporate graph/model uncertainty, and extend the framework from single node/link analysis to concurrent case where criticality scores would be assigned to a group of nodes/links. Then, the task would be to identify a critical set of nodes/links, and this could find potential applications in attack graphs.

\section*{Acknowledgement}
This material is based upon work supported by National Science Foundation under award number $1855216$.

\bibliography{reference}

\end{document}